\documentclass[12pt]{iopart}
\usepackage{iopams}
\usepackage{epsfig}

\newcommand\ie{\textit{i.e.}}
\newcommand\eg{\textit{e.g.}}

\newcommand\eqref[1]{(\ref{#1})}

\newcommand\smallkern{\kern0.1em}
\newcommand\wall{\unkern\smallkern|\smallkern}
\newcommand\nonmag{\unkern\smallkern|\kern-0.1em|\smallkern}

\newcommand\bkt[1]{\langle #1\rangle}

\begin{document}
\clubpenalty1000
\widowpenalty1000
\catcode`@11
\let\c@footnote\relax
\catcode`@12
\newcounter{footnote}[page]

\title[Multiphase region of helimagnetic superlattices]%
	{Multiphase region of helimagnetic superlattices
	at low temperature in an extended six-state clock model}
\author{D C Lovelady, H M Harper, I E Brodsky,\footnote{%
Current address: Department of Mathematics, University of Florida,
Gainesville, FL, USA}
and D A Rabson\footnote{Corresponding author}}
\address{Department of Physics, PHY 114, University of South Florida,
4202 East Fowler Avenue, Tampa, FL 33620, USA}
\ead{davidra@ewald.cas.usf.edu}

\begin{abstract}
The variety of magnetic phases observed in rare-earth
heterostructures at low temperatures \cite{Jehan}, such as Ho/Y,
may be elucidated by an ANNNI-like model Hamiltonian.  In
previous work modelling bulk Ho \cite{Seno}, such a Hamiltonian
with a one-dimensional parameter space produced a single
multiphase point.  In contrast, the parameter space of the
heterostructure model is three-dimensional, and instead of an
isolated multiphase point, we find two-dimensional multiphase
regions.  In an example of Villain's ``order from disorder''
\cite{Villain80a,Villain91}, an infinitesimal temperature breaks
the ground-state degeneracy.  In first order of a
low-temperature expansion, we find that the degeneracy is broken
everywhere in a multiphase region except on a line.  A segment
of the line appears to remain multiphase to all orders in a
low-temperature expansion when the number $L$ of magnetic layers
between non-magnetic spacers is 4 but not for other values of
$L$.  For $L=4$, the hierarchy of phases more closely resembles
that in the ANNNI model than in the bulk six-state clock model
on which the present model is based.
\end{abstract}
\pacs{64.60.Cn, 05.50.+q, 75.70.Cn}
\vspace{28pt plus 10pt minus 18pt}
\noindent{\small\it Published version: J.\ Phys.\ A: Math.\ Gen.\ \bf39\rm (2006) 5681--5694}

\section{Introduction}\label{sec:intro}

Layered planes of rare-earth
metals exhibit a wealth of magnetically-ordered phases
at low temperature.  In helimagnetic phases,
spins (treated classically) align ferromagnetically within
each plane, with an axial RKKY interaction responsible for a
progression of spin angles through successive
planes \cite{Blundell, Jensen}.
Strong easy-axis anisotropy may
frustrate the natural RKKY pitch angle, leading to a
multitude of possible phases characterized by the number of layers
separating skips, or ``walls,'' in the pattern of pitch angles.
In the axial-next-nearest-neighbour
Ising (ANNNI) \cite{Fisher80,Fisher81,Szpilka86,Fisher87} and
related clock models \cite{Yeomans3,Yeomans2,Sasaki92,Seno,Pleimling,Neubert98},
a single parameter
controls the relative strengths of competing interactions, and at
a single value of this parameter, infinitely many phases coexist;
this is called a multiphase point.
Since these phases cover
all allowed spacings between walls,
such phases are indistinguishable
from random sequences.  Thus the zero-temperature state is disordered.  This
disorder is broken at infinitesimal temperature in an example of ``order
from disorder'' \cite{Villain80a,Villain91}.  We now ask what happens 
in a model of helimagnetic heterostructures with a three-dimensional
parameter space: we identify
fully two-dimensional multiphase regions and investigate the
topology of the low-temperature phase diagram.

With the giant magnetoresistive effect \cite{GMR88} in
ferromagnetic/nonmagnetic superlattices having spawned
important technological applications that reached the market
around 1997 \cite{Derbyshire95,Araki00}, it seems
practical, as well as theoretically interesting, to examine
the possible phases of
helimagnetic/nonmagnetic superlattices.
Such superlattices have been deposited using molecular-beam
epitaxy, alternating dysprosium \cite{Erwin}, erbium \cite{Borchers},
or holmium \cite{Jehan} with non-magnetic yttrium spacer layers
as well as holmium with lutetium \cite{Swaddling96}.
Surprisingly, neutron-scattering experiments show that the helicity 
of the spins in the rare-earth layers is preserved across the spacers,
with the magnetic moments forming long-period ``spin-slip''
phases \cite{Jehan}.
RKKY-like polarization \cite{Bruno95} of conduction electrons in the non-magnetic
layers is again implicated
\cite{Cowley98,Cowley99}; in any case, we
can model the indirect exchange across non-magnetic spacers in parallel with
that between successive magnetic planes.  If the exchange parameters
can be controlled with pressure, external fields,
or spacer-layer thickness, such
systems could possibly be useful as magnetic sensors or in data-storage
applications.  Axially modulated, high-order, commensurate phases
are not limited to rare-earth heterostructures:
Szpilka and Fisher \cite{Szpilka86} cite half a dozen
other systems in which such
phases have been observed,
ranging from CeSb \cite{RossatMignod80} to ferroelectric
thiourea \cite{Moudden83,Durand84}.

Seno \textit{et al.} \cite{Seno} applied the ANNNI ideas to
a case of infinite hexagonal anisotropy, the six-state clock
model, relevant, for example, to bulk holmium.\footnote{%
An extension of this work
presented a small-inverse-anisotropy expansion about the clock
model and again found a hierarchy of phases emanating
from the multiphase point at infinite anisotropy \cite{Seno94}.}
A spin $\alpha$ in the $j^{\rm th}$ plane
points in a direction that is an integral multiple,
$n_{j\alpha}$, of $2\pi/6$.
At zero temperature, all the spins in a plane point
in the same direction ($n_j$), and the model is controlled
by a single parameter, the ratio $x$ of the strength of the
next-nearest-axial-neighbour antiferromagnetic ($J_2$)
to nearest-axial-neighbour
ferromagnetic ($J_1$) interaction,
with the axial 
terms in the Hamiltonian summing
$-J_1\cos(2\pi(n_{j+1,\alpha}-n_{j,\alpha})/6)$
and $+J_2\cos(2\pi(n_{j+2,\alpha}-n_{j,\alpha})/6)$.
For $0<x<1/3$, the ground state is a ferromagnet, for $1/3<x<1$ a helimagnet
with no walls, and for $x>1$ a helimagnet interrupted by walls every second
layer.  At the single point $x=1$ in the one-dimensional phase diagram,
infinitely many phases coexist in the ground state.
We represent the
helimagnetic phase ($1/3<x\le1$) by the axial sequence $\dots012345012\dots,$
understanding that this includes as well the translations and reflections
of the sequence.  The two coexisting period-2 phases for $x>1$ are represented
by $\dots00330033\dots$ and $\dots01\wall34\wall01\wall34\dots$:
this last is thought of as a modification of the helical phase by the
insertion of skips, or walls (denoted ``$\wall$''), every second layer.
The walls are analogous to domain walls in the ANNNI model.
At the multiphase point, $x=1$, in addition to $\dots00330033\dots,$
a helical phase with walls placed anywhere at least two layers apart
is a ground state of the system, \eg,
$\dots01\wall345\wall12\wall450\dots$.  A convenient notation
in ANNNI-type models labels a periodic phase by the spacings between
successive walls: thus, this last example is $\bkt{23}$, the
phase with walls every second layer $\bkt{2}$, and the bare helical
phase without walls $\bkt{\infty}$.  In a low-temperature expansion,
Seno \textit{et al.} followed a hierarchy of phases (similar to what
we describe below)
and showed that
each phase between $\bkt{23}$ and $\bkt{\infty}$ acquires a region
of stability at infinitesimal temperature.

The forgoing model simplifies the actual magnetic structure
of bulk holmium.   Neutron scattering gives
the turn angle per atomic layer as
$30^\circ$ rather than $60^\circ$,
with moments bunched
in pairs around the six easy axes \cite{Bohr,Jehan}, and
while the average turn angle increases in films, the effect is thought to be
due to interspersal of singlets among the pairs; thus the $\bkt3$
phase in the simplified model might actually represent
moments $\dots00122344\dots$, where pairs of repeated spins
lie a few degrees before and after the easy-axis direction (see Fig.~14
of Reference \cite{Jehan}).
The model, or its present extension to superlattices, was meant not
to reproduce realistic details of a particular rare-earth helimagnet
but rather to reduce a system with competing crystal-field and exchange
interactions to the simplest form, in which exact results
are possible, so as to investigate universal properties of the
resulting hierarchy of commensurate, longitudinally-modulated
spin-slip phases.

\section{The model and its ground states}\label{sec:ground}
We consider a superlattice in which blocks of $L$ magnetic
layers are separated by non-magnetic spacers
characterized by effective couplings $J_1'$ and $J_2'$;
this simple extension of the bulk model of \cite{Seno}
gives the full Hamiltonian
\begin{equation}
\eqalign{
\fl
\mathcal{H} ~=~ -{1 \over 2}J_{0}\sum_{i,\alpha,\beta(\alpha)}\cos\left({2\pi 
\over 6}(n_{i\alpha}-n_{i\beta})\right)\cr
-J_{1}\sum_{i,\alpha}\cos\left({2\pi \over 6}(n_{i\alpha}-n_{i+1,\alpha})\right)
+J_{2}\sum_{i,\alpha}\cos\left({2\pi \over 6}(n_{i\alpha}-n_{i+2,\alpha})\right)
\cr
-J_1^{\prime}\sum_{i,\alpha}\vphantom{\Sigma}'
\cos\left({2\pi \over 6}(n_{i\alpha}-n_{i+1,\alpha})
\right)
+J_2^{\prime}\sum_{i,\alpha}\vphantom{\Sigma}'
\cos\left({2\pi \over 6}(n_{i\alpha}-n_{i+2,\alpha})
\right)\rlap{\quad,}
\cr
}
\label{eqn:ham}
\end{equation}
where $i$ labels layers, $\alpha$ a spin
within a (simple-hexagonal) layer, and $\beta(\alpha)$ its nearest neighbours.
The unprimed sums in the second line are taken only over
bonds that do not straddle a non-magnetic spacer, while the primed sums in
the third line are taken only over bonds that do.  For purposes of the
low-temperature expansion, the in-plane ferromagnetic coupling constant
$J_0$ is taken to be positive and much stronger than any of the axial couplings
\cite{Seno}.
Since we are looking for helical phases, we take
all of the remaining four couplings also to be positive.  (Certain
negative couplings are in fact related to the positive sector by
symmetries of $\mathcal{H}$.)
The model reduces to that of \cite{Seno}
when $J_1'=J_1$ and $J_2'=J_2$ or, equivalently, when $L=1$.
The three-dimensional coupling space
is given by $x=J_2/J_1$, $y=J_1'/J_1$, and $z=J_2'/J_1$; it is convenient
to set $J_1=1$.

We generalize the previous notation to accommodate states of
a superstructure in
which blocks of $L$ magnetic layers are separated by non-magnetic
spacers, denoted by $\nonmag$, with the arrangement repeated periodically.
(The symbol $\nonmag$ may denote any number of atomic layers of the
non-magnetic metal.)
Since the direct interactions in \eqref{eqn:ham}
extend a maximum of two layers in the axial direction,
walls are classified in three categories.  A wall at least two layers
from a spacer has the same energy cost as in the bulk
model and is termed a type-1 wall, for example ($L=5$)
\begin{equation}
\dots\nonmag0123\wall50\nonmag12\dots\rlap{\quad.}
\label{eqn:type1}
\end{equation}
Insertion of a wall one layer from a non-magnetic spacer has a different
energy cost, since a $J_2'$ bond is broken.  This is termed a type-2 wall:
\begin{equation}
\dots\nonmag01234\wall0\nonmag12\dots\rlap{\quad.}
\label{eqn:type2}
\end{equation}
A type-3 wall coincides with a non-magnetic spacer:
\begin{equation}
\dots\nonmag012345\nonmag\wall12\dots\rlap{\quad.}
\label{eqn:type3}
\end{equation}
Helical configurations, including $\bkt\infty$ itself,
that differ from $\bkt\infty$ only by the insertion of
walls are called wall states.  These states preserve the
sense of helicity (positive or negative).
We consider $L\ge3$, as $L=1$ is the same as bulk,
while $L=2$ omits the $J_2$ ($x$) parameter
and so has only a two-dimensional parameter space.  It is also
less likely to be of experimental interest.

\begin{figure}[htb]
\centerline{\epsfxsize0.6\hsize\epsfbox{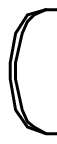}}
\caption{\label{fig:yz}
Ground-state phase diagram for (part of) the $y=z$ plane, $L=4$.
The horizontal axis gives the normalized bulk second-neighbour coupling,
the vertical the couplings across non-magnetic spacers.  Outside the
triangle delimited by dot-dashed lines, the ground states are
as indicated.  Inside the triangle, wall states are the ground states.
(An exhaustive search found no lower-energy states of length up to $3L$.)
Similar ground-state phase diagrams were calculated for other values of $L$.
The first-order low-temperature expansion gives $\bkt\infty$
to the left of the dotted line within the triangle and the $\bkt1$
phase to the right; on the
line itself, these phases and their progeny coexist, requiring a
higher-order low-temperature expansion to distinguish.  On roughly
the upper half of the left leg of the dotted line, from $z=11/9$ to
$z=13/9$, we believe infinitely many phases of the form
$\bkt{1^k2}$ coexist to all orders.
}
\end{figure}

A straightforward calculation yields the total energy of a wall state
as a function of the densities $W_i$ of walls of the the three types:
\begin{equation}
\fl
\eqalign{
E_{\rm wall} ~=~\cr
\quad-{1 \over 2L}\Bigl((1+x)(L-2)+1+y+2z\Bigr) +
(1-x)W_1 +
\Bigl(1-{x+z \over 2}\Bigr)W_2 +
(y-z)W_3\rlap{\quad.}\cr
}
\label{eqn:ewall}
\end{equation}
As in the original model, successive walls are energetically forbidden.
We seek regions of the three-dimensional parameter space in which
the insertion of a wall of some type costs no energy: this occurs
when the coefficient of one or more of the densities $W_i$
vanishes.  Thus the planes $x=1$, $(x+z)/2 = 1$, and $y=z$ all
potentially constitute multiphase regions; however, it is also
necessary to consider competing non-wall states, which may have
lower energies.  For present purposes, we shall concentrate
on the $y=z$ plane, for which type-3 walls cost no energy.
Since a negative energy for type-2 walls would shut type-3 walls
out, we examine the part of the $y=z$ plane to the left of
the $x+z=2$ line.  By considering points to the left of the line
$x=1$, we exclude type-1 walls as well.
For $L=4$, direct calculation gives the
phase diagram of Figure~\ref{fig:yz}.
An exhaustive computer search (of phases of length $3L=12$
with twisted periodic boundary conditions) verified that
the wall-state energy \eqref{eqn:ewall} is lower than that of any
competing phase inside a triangle in the $y=z$ plane, which constitutes a
multiphase region.  Comparable wall-state regions were calculated numerically
for $L$ in the range 3--11.

Since we are concentrating on a region in which type-1 and type-2 walls
are excluded, while type-3 walls cost no energy, we adapt the
notation of \cite{Seno} to count magnetic blocks, rather than magnetic
layers, between walls.  Thus, for example, with $L=4$, $\bkt1$
has a wall coinciding with each spacer, while the $\bkt2$ phase has a wall at
every other spacer.  Since no restriction prevents adjacent walls of this
type, the
count of possible phases is simply $2$ to the power of the number of
magnetic blocks; this represents a simplification relative to the
ANNNI and other related models \cite{Redner81}.

\section{Low-temperature expansion: first order}\label{sec:lowT1}
The novel feature presented by the current problem is the multiphase
triangle (for $L=4$ or a similar polygon for other $L$) throughout which
infinitely many phases coexist at zero temperature.  An interesting
theoretical question is how thermal disorder can distinguish the
free energies of all these phases in the given region.

Although the Hamiltonian \eqref{eqn:ham} contains only first-
and second-neighbour axial terms, a non-zero temperature introduces
effective long-range interactions through an axial chain of
thermally-excited spins, each pointing in a direction at variance
with its in-plane neighbours \cite{Fisher80,Fisher81,Seno}.  By
analogy to the ANNNI model, we call such excitations ``spin flips.''
Since
the number of ways an excitation of a particular energy may
occur depends on the state, flipped spins provide an entropic
mechanism for distinguishing the free energies of wall states at
infinitesimal temperature.  If the $i^{\rm th}$ excitation, which
may involve several spins, has an energy $\Delta E_i$ relative to the
ground-state energy per spin $E_0$ and can be placed on the lattice
of $N$ spins
$g_i$ different ways, the free energy per spin is given by
the linked-cluster theorem \cite{Wortis}:
\begin{equation}
f = E_0 - k_B T \sum_i \gamma_i e^{-\beta\,\Delta E_i}\rlap{\quad,}
\label{eqn:linkedcluster}
\end{equation}
where $\gamma_i=\lim_{N\rightarrow0}g_i/N$
is the intensive part of $g_i/N$.  (The limit discards
those terms in $g_i$ that go as higher powers of $N$;
such terms come from independent clusters of spin excitations.)

We apply the method first to an isolated spin flip, which may occur
in a layer adjacent to or one layer separated from a spacer, or it
may ($L>4$) occur in bulk.  An isolated spin flip in bulk gives the
same contribution to $f$ regardless of phase, so we calculate the
energies and counts $\gamma_i$ just for the first two cases, leading to
the weights in Table \ref{tab:firstorder}.
The case $L=3$ requires special treatment because the cost of
an excitation in the layer in the middle of a block depends on
the presence or absence of walls on both sides.

\begin{table}[htb]
\caption{\label{tab:firstorder}Contributions to \eqref{eqn:linkedcluster}
are formed by a count (per spin) of the number of ways of forming the
excitation times a Boltzmann factor.  The left column gives an
example of the excitation under consideration, where the caret ($\wedge$)
marks the plane in which a single spin is rotated (``flipped'')
plus $60^\circ$ or minus $60^\circ$
from the angle of its neighbours in the plane.
The second column gives the Boltzmann factor, and the remaining
columns give the intensive counts $\gamma_i$ weighting the Boltzmann
factor for the cases $\bkt1$, $\bkt2$, and $\bkt\infty$.
$L$ is the number of magnetic layers in a block.
The last three rows apply only to $L=3$.
Here, $\beta$ is the inverse
temperature, $q=\exp(-\beta J_0/2)$,
$t\,[=6]$ the number of in-plane nearest neighbours, and
$r=\exp(-\beta J_1/2)$.}
$$
\vcenter{
\everycr{\noalign{\vskip0.5\baselineskip}}
\halign{\tabskip2em#\hfil&$#$\hfil&$\displaystyle#$\hfil%
&\hfil$#$\hfil&\hfil$#$\hfil&$#$\hfil\tabskip0pt\cr
\ns\ns\br
&&&\multispan3{\hfill\textrm{intensive count}\hfill}\cr\ns
&\textrm{excitation} & \textrm{Boltzmann factor}%
&\bkt1&\bkt2&\bkt\infty\cr
\ns\mr
1.&45\hat01\nonmag23 & \,q^t\big(r^{1-x+2z} + r^{1+2x-z}\big) &%
0 & 1/L &  2/L\cr
2.&450\hat1\nonmag23 &%
\,q^t\big(r^{2-x-y+2z} + r^{-1+2x+2y-z}\big)&%
0 & 1/L & 2/L\cr
3.&45\hat01\nonmag\wall34 & q^t\big(r^{1-x+z} + r^{1+2x+z}\big)&%
2/L & 1/L & 0\cr
4.&450\hat1\nonmag\wall34 & q^t\big(r^{2-x-2y+z} + r^{-1+2x+y+z}\big)&%
2/L & 1/L & 0\cr
5.&0\nonmag1\hat23\nonmag4 & 2 q^t r^{1+z} &%
0 & 0 & 1/L\cr
6.&0\nonmag\wall2\hat34\nonmag5 & q^t\big(r+r^{1+3z}\big)&%
0 & 1/L & 0\cr
7.&0\nonmag\wall2\hat34\nonmag\wall0 & 2q^t r^{1+2z}&%
1/L & 0 & 0\cr
}
}
$$
\end{table}

We consider $L\ge4$ first.  If there are no type-3 walls,
the only single-spin excitations (other than bulk) will be of one of the
types in the first two rows of Table \ref{tab:firstorder}.  This describes
the $\bkt\infty$ phase.  If a phase
has the maximum density of type-3 walls, the excitations
will be of the types in the second two rows.  This is the $\bkt1$ phase.
To this first order in the low-temperature expansion, any other wall
phase (\eg, $\bkt2$)
will have a free energy intermediate between these two cases.
Thus we look first for the coexistence of $\bkt1$ and $\bkt\infty$.
Subtracting rows 1 and 2 from the sum of rows 3 and 4 gives the
free-energy difference
\begin{equation}
\eqalign{
\Delta f ~&=~ f_{\bkt1} - f_{\bkt\infty}\cr
~&=~ -{2\over L}k_B T q^t\,
\left( r^{1-x+z} + r^{1+2x+z} + r^{2-x-2y+z} + r^{-1+2x+y+z} \right.\cr
&\quad\left.\quad - r^{1-x+2z} - r^{1+2x-z} - r^{2-x-y+2z} -r^{-1+2x+2y-z} \right)
\rlap{\quad,}\cr
}
\label{eqn:deltaf}
\end{equation}
where $\beta=1/(k_B T)$ is the inverse temperature, $t$ the number
of in-plane nearest neighbours, $q=\exp(-\beta J_0/2)$,
and $r=\exp(-\beta J_1/2)$.
Setting $\Delta f=0$ and $y=z$ yields the expression
\begin{equation}
r^{3x} = {{r^z + r - 1 - r^{1-2z}}\over{1 + r^{z-2} -r^{-2z} - r^{-2}}}
\rlap{\quad.}
\label{eqn:r3x}
\end{equation}
In the zero-temperature limit, $r\rightarrow0$, so the power of $r$
with the smallest exponent dominates.  This allow us to
solve for the coexistence line,
\begin{equation}
x = \cases{{2\over 3}&for $0<z\le{1\over 2}$\cr
        1-{2\over 3}z&for ${1\over2}\le z\le1$\cr
        {1\over 3}&for $1\le z\le2$\cr
}\rlap{\quad\quad\hbox{$(L\ge4)$},}
\label{eqn:zigzag}
\end{equation}
drawn as a dotted line in Figure~\ref{fig:yz}.  In the multiphase region
to the left of this line, the $\bkt\infty$ phase has the lowest free energy,
breaking the infinite degeneracy of zero temperature.\footnote{%
The twelvefold degeneracy of $\bkt\infty$ neither scales with
$N$ nor affects the spin-spin correlation function.}
To the right of the line, the $\bkt1$ phase dominates.  On the line
itself, all wall phases remain degenerate; to break the degeneracy
it will be necessary to consider more flipped spins.

First, however, the model with $L=3$
introduces a new element to the low-temperature expansion.  In the last
three rows of Table~\ref{tab:firstorder}, the count of the $\bkt2$ phase 
does not merely interpolate between the counts of $\bkt1$ and $\bkt\infty$:
that is, a single-spin excitation in the middle plane of a magnetic block
distinguishes not only $\bkt1$ from $\bkt\infty$ but also
each from $\bkt2$.  Thus, the first-order expansion must potentially
consider three coexistence lines.  In the event, the three collapse
to one.  For $z>0$, all wall phases coexist on the line
\begin{equation}
z = \frac32(1-x)\rlap{\quad\quad\hbox{($L=3$)}.}
\label{eqn:zag3}
\end{equation}
For $z>(3/2)(1-x)$, the $\bkt1$ phase has the lowest
free energy, while for smaller $z$, the $\bkt\infty$ phase has
the lowest free energy.

\section{Expansion to higher orders}\label{sec:lowT+}
The hierarchy of potential phases in the low-temperature
expansion has been described well elsewhere
\cite{Fisher80,Fisher81,Yeomans2,Yeomans1}
and so will only be summarized.  At any order of the expansion,
a coexistence region has been established between two ``parent''
phases and infinitely many other wall states.  (In Figure~\ref{fig:yz}
for $L=4$, this region is the zig-zag line, on which,
to first order, parents $\bkt1$
and $\bkt\infty$ coexist with all other wall states.)  Spin excitations
to this order of the expansion cannot distinguish the parents from
the other wall states, but by adding some number of additional spin
excitations, linked to those of the given order, we can distinguish
the two parents from a ``child'' phase made by concatenating one
period of each parent.  As examples, the child of $\bkt1$ and $\bkt\infty$
is $\bkt2$, while that of $\bkt1$ and $\bkt2$ is $\bkt{12}$.
A connected chain of spin excitations
can ``see'' the presence or absence of walls over its length;
viewed another way, this leads to an effective long-range interaction
between walls.

While in principle one could continue the enumeration of connected
excitations of two, three, and more spins along the lines of
Table~\ref{tab:firstorder},
a transfer-matrix technique \cite{Yeomans3,Seno} is well
suited to computer symbolic algebra.  We defer
implementation details to the appendix.  The matrices are
more involved than those in \cite{Seno},
so the results are for specific cases, from
which we conjecture generalizations.

In first order, we have already seen the two-dimensional multiphase
region shrink to one dimension (Figure~\ref{fig:yz}).  We wish
to find out whether the line shrinks further to a point or set of points,
or whether the line, or a portion of the line, behaves like a
multiphase point, with the additional degree of freedom essentially
irrelevant.  It is also of interest whether all wall states descending
from $\bkt1$ and $\bkt\infty$ attain stability or only a subset.

We carried out the low-temperature expansion for
magnetic blocks of length $L$ between $3$ and $17$;
except for the interesting case of $L=4$, the
hierarchy terminates after just a few phases.
Aside from $\bkt\infty$, the only stable phases found for $L=4$ were
of the form $\bkt{1^k2}$, $0\le k\le27$ (the highest calculated)
and $k=\infty$ (\ie, $\bkt1$).
This resembles the ANNNI model \cite{Fisher80,Fisher81} more than
some clock models
in that there do not
exist two phases\footnote{$\bkt{23}$ and $\bkt{\infty}$ in the bulk
model \cite{Seno}}
\textit{all} of whose progeny attain stability.
Villain and Gordon \cite{Villain80b} (see also \cite{Szpilka86})
distinguish a Devil's staircase
\cite{Bak80} from a ``harmless'' one.  In both, a multiphase
point gives rise to a large number of phases that approaches infinity
at $T\rightarrow0$.  However, in the latter case, at any \textit{finite}
$T>0$, it is argued that only finitely many phases are stable.  Since
our model fails to find an infinite hierarchy of ``mixed phases''
\cite{Yeomans2}, we conjecture that our staircase may similarly be
harmless.

The way the $\bkt1$-$\bkt\infty$ coexistence line breaks
up for $L=4$ is also of interest.  It intersects the multiphase
triangle (Figure~\ref{fig:yz}) for $1/3\le z\le13/9$; outside
this region, it ceases to describe coexistence of \textit{ground} states.
The symbolic transfer-matrix calculation finds that $\bkt{2}$
is stable on the line only 
for $3/4\le z\le13/9$.  Below $3/4$, there is a first-order
phase transition between $\bkt1$ and $\bkt\infty$.
The phase $\bkt{12}$ is stable
at $z=3/4$ and then again for $11/9 \le z \le 13/9$.  All subsequent
phases $\bkt{1^k2}$ for which we were able to extract symbolic results
($k\le5$)
are stable for $11/9<z\le13/9$ (that is, $3/4$ and $11/9$ drop out).
Numerically, $k=$6--27 is stable for $11/9\lesssim z\le13/9$, the ``$\lesssim$''
indicating the inability of the numerical code to distinguish between
the proper and improper inequality.  We conjecture that $\bkt{1^k2}$ is
stable for $11/9\le z\le13/9$ for all $k\ge2$ and that ``mixed phases''
never come in, something we were able to confirm up to the mixed phase
$\bkt{1^{18}\,2\,1^{17}\,2}$.

For $L=3$, the coexistence line \eqref{eqn:zag3} intersects with
the region in which wall states have the lowest energy for $0\le z\le1$.
The states $\bkt1$, $\bkt\infty$, $\bkt2$, $\bkt{12}$, and $\bkt3$
are stable on this line segment, but no other phases.

For $L=5$, the coexistence line is again \eqref{eqn:zigzag}, which
passes through the wall-state region for $0\le z\le13/9$.   The same
phases are stable as for $L=3$: $\bkt2$ for $0<z\le3/4$, $\bkt{12}$
for $0<z\le3/4$, and $\bkt3$ for $0<z<3/4$.  For $L=6$, the
$\bkt1$-$\bkt\infty$ coexistence line gives the lowest energy for
$0\le z<13/9$; however, the only other stable phase is $\bkt2$,
and only at the single point $z=3/4$.

The following pattern appears to hold for $L>6$: the
coexistence line \eqref{eqn:zigzag} intersects with the wall-state region
for $0\le z\le13/9$.
Even values for $L$ (we
computed 8, 10, 12,
14, and 16) give a first-order transition between $\bkt1$ and $\bkt\infty$
all along the coexistence line.  No other phases are stable.  For odd
$L$ (7, 9, 11, 13, 15, and 17), the phases $\bkt2$, $\bkt{12}$,
and $\bkt3$ are also stable
for $0<z<3/4$.

\section{Implications}\label{sec:implications}
\begin{figure}[htb]
\centerline{\epsfxsize0.6\hsize\epsfbox{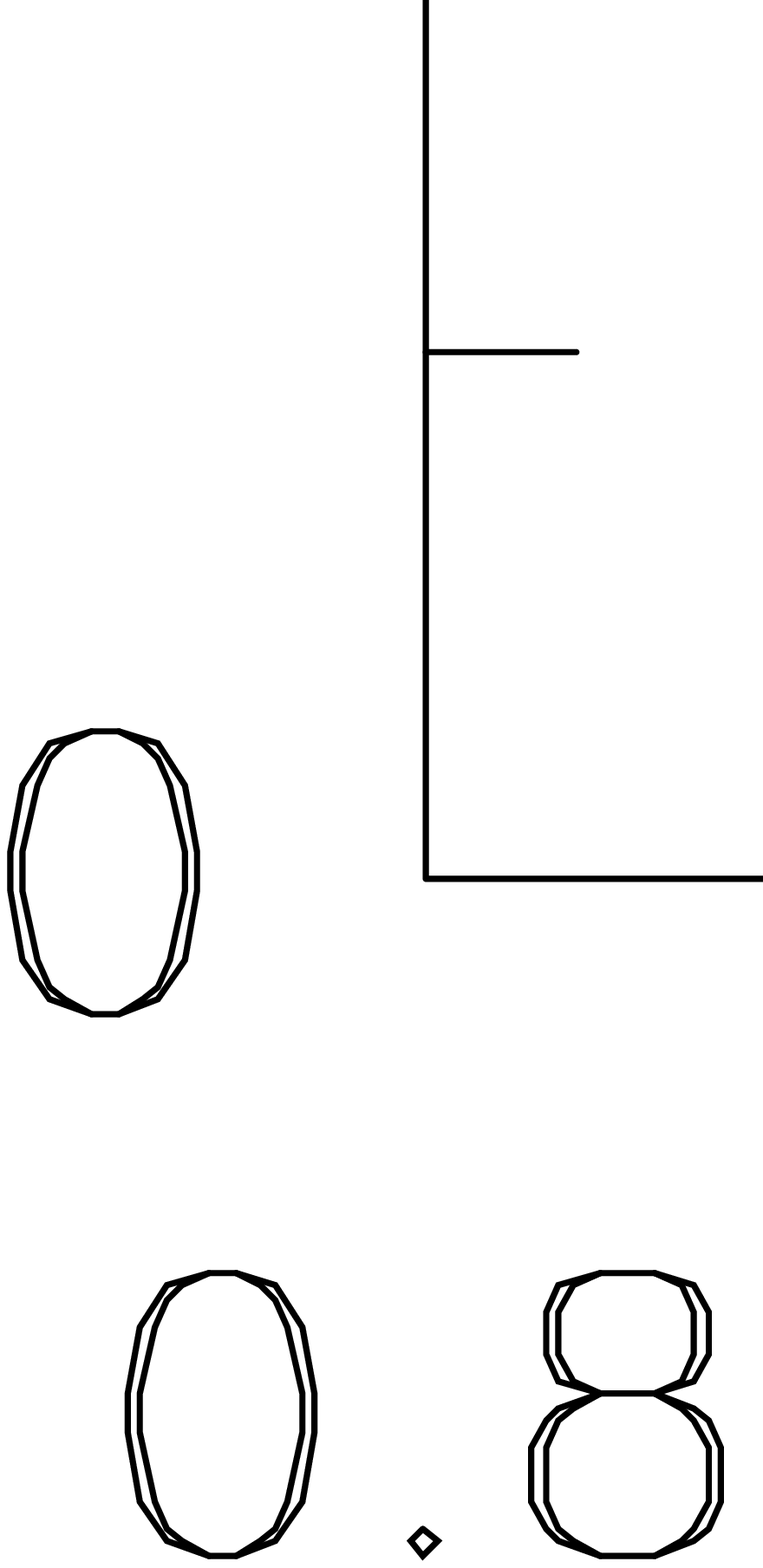}}
\caption{\label{fig:mft}
Diagram of a hierarchy of phases
emanating from a multiphase point at
zero temperature up to a Curie temperature.
The horizontal axis represents a ratio $x$ of coupling strengths,
which at $1$ leads to zero-temperature disorder.  Raising the
temperature from the vicinity of $x=1$ gives a succession of
stable phases.
Adapted from Figure~1 of \cite{Seno}, where it shows
a numerical mean-field calculation on the bulk six-state clock model.
In the present context, it can be thought of as schematic for the
$T>0$ behaviour of the system of Figure~\ref{fig:yz} at some point along the
zig-zag line, where $x$ represents a transverse dimension.
}
\end{figure}

The low-temperature expansion applies at infinitesimal temperature,
but the bulk model has also been investigated with a mean-field
theory, which should be valid only at high temperature \cite{Seno}.
The low-order phases predicted by the low-temperature
expansion were seen to spread out from the multiphase point as
temperature increased (Figure~\ref{fig:mft});
the only notable discrepancy between the
two extreme theories was the presence of phases $\bkt{2^k3}$ in the
mean-field calculation, and this was explained in terms of a competing phase.
Near the zero-temperature multiphase point, which in the $L=4$
model would be replaced by the multiphase zig-zag line of Figure~\ref{fig:yz},
the spin-spin correlation length is expected to be small, as
a large number of commensurate phases with different periods coexist.
At any temperature greater than zero and less than the Curie temperature,
only one phase is stable; however, in an experimental system, interfacial
roughness and interdiffusion
might lead to coexisting commensurate phases from nearby
points in the temperature-phase space.
As the temperature increases, the volumes
of stability do as well, so that no phases lie nearby, thus stabilizing
a single phase.

Interestingly, the coherence lengths $\xi$ of the basal-plane holmium
moments in Ho/Er
superlattices have been found to {\it increase\/} with
temperature $T$ between $8$K to $100$K \cite{Simpson94}.
Since Er acquires a moment below
$100$K, the experimental system is considerably more complex than
our simple model; moreover, too few temperatures were measured
to permit a comparison to the plateaux one would expect in $\xi(T)$
from Figure~\ref{fig:mft}.  A similar effect is observed in
Er/Lu \cite{Simpson97,Cowley98}.

The question of commensurate {\it versus\/} incommensurate magnetic
modulation also awaits experimental resolution.  In the low-temperature
expansion, incommensurate phases are only approached, as the limit
of a hierarchy of commensurate phases, while the bulk mean-field calculation
(Figure~\ref{fig:mft}) suggests that these limiting phases will occupy
a volume of measure zero in the phase diagram.
While several rare-earth systems unambiguously show commensurate phases
\cite{Bohr,Borchers,Simpson97}, other superlattices appear to show
a continuous increase with temperature
in the average turn angle per atomic layer, suggesting that
incommensurate phases are generic \cite{Jehan,Cowley04}.
We cannot rule out an averaging effect being responsible, but this
would appear inconsistent with the absence of plateaux and
the expectation of vanishing measure for high-order phases.
It will
be particularly interesting to investigate whether a statistical-mechanical
model not much more complex than that considered here can incorporate
more of the qualitative behaviour seen in rare-earth superlattices.

We have shown that
a superlattice of helimagnetic and non-magnetic layers
exhibits behaviour different from that of the bulk six-state clock
model \cite{Seno}.  There are multiphase regions, rather than
a single multiphase point.  When precisely four magnetic layers
lie between non-magnetic spacers, a line segment in the multiphase
triangle appears to support a set of phases more like that in the
ANNNI model \cite{Fisher80,Fisher81} than like the bulk six-state
clock model.
For other values of $L$, the low-temperature expansion finds only
a few stable phases.  This raises the interesting experimental
question of whether rich magnetic phase diagrams in artificial
superlattices could appear for certain magic spacings while being
absent for others.  If the phase diagram were to depend as
sensitively on $L$ as in our model, it might be difficult to
grow films sufficiently uniform to test the hypothesis; however,
if the extent of the magic coupling were broader (say, L=4--6),
the effect could be observable.  Further, a
multiphase region of coupling space might be
more amenable to experiment than a multiphase point that requires
exact tuning; such a region, however, would need to have the full
dimensionality of the coupling space, something we have not yet
constructed.

\section*{Note}
In the published version of this paper, the reference at the
end of the first sentence of Section~\ref{sec:implications} reads [12].
It should be \cite{Seno}.

\ack{DAR is a Cottrell scholar of Research Corporation, which
has supported this research.  We also acknowledge Roxane Rokicki's assistance
at an early stage of this work.  Calculations were performed at
the Research Computing Core Facility at the University of South Florida.}

{
\appendix
\section*{Appendix}
\setcounter{section}{1}
In order to calculate the free-energy difference of a child from its parents,
we adapt the transfer-matrix technique \cite{Yeomans3,Fisher87,Seno}
to the region with only type-3 walls.
We begin in a region over which, to the order already calculated in
the low-temperature expansion, parent phases
$\bkt{a}=\bkt{a_1\,a_2\,\dots}$ of period $p_a=\sum_i a_i$ and
free energy per spin $f_a$ and
$\bkt{b}=\bkt{b_1\,b_2\,\dots}$ of period $p_b=\sum_i b_i$ and
free energy per spin $f_b$
coexist and
have lower free energies than their parent phases.\footnote{%
The period of $\bkt\infty$, for the purpose of \eqref{eqn:energydiff}, is $1$.}
We then seek the double free-energy difference
\begin{equation}
a_{\bkt{ab}} = f_{\bkt{ab}} - {{p_a}\over{p_a+p_b}} f_{\bkt{a}}
-{{p_b}\over{p_a+p_b}} f_{\bkt{b}}
\label{eqn:energydiff}
\end{equation}
to leading order.  If $a_{\bkt{ab}}<0$,
the child phase $\bkt{ab}$ acquires a region of stability.  Isolated spin
rotations (as in Table \ref{tab:firstorder}) cannot determine the
sign of \eqref{eqn:energydiff}, since the three phases, $\bkt{a}$,
$\bkt{b}$, and $\bkt{ab}$, have the same free energies to first order.
We must consider connected spin excitations:
in general, the Boltzmann weight of two (or more) spin rotations that
share an axial bond will differ from the weight of the same rotations
situated in their respective planes such that they do not share
a bond.  Since the $J_0$ (in-plane)
bond is assumed the most expensive to break,
the shortest excitation that distinguishes $\bkt{ab}$ from its
parents provides the leading term in the low-temperature expansion.
This requires that the connected excitation should span
$(p_a+p_b-1)$ blocks of length $L$, in the sense that bonds on
each end extend through the terminating spacer layers and so
sense whether these spacers coincide with walls.
The transfer-matrix technique keeps track of all the combinations of
connected and disconnected excitations of this length.

As in \cite{Seno}, two cases arise.  When the product
$(p_a+p_b-1)\cdot L$ is odd, an excitation of connected spins every
second layer
distinguishes the child from the two
parents, and $2\times2$ matrices suffice.
When the product is even, we shall need $4\times4$ matrices.

The principles are best illustrated
by an example.
Consider distinguishing $\bkt2$ from its parents $\bkt\infty$ and
$\bkt1$ when $L=5$. In the following diagram showing just over
one period of the $\bkt2$ phase, $S$ represent a magnetic layer,
while $\hat S$ represents a magnetic layer with a flipped spin:
\begin{equation}
S\nonmag \wall S \hat S S \hat S S \nonmag S S S S S \nonmag \wall S
\rlap{\quad.}
\label{eqn:2phase}
\end{equation}
In the $\bkt2$ phase, the
two extremal spacers ($\nonmag$)
coincide with walls.  In the $\bkt1$-phase parent,
all three spacers coincide with walls, while in the $\bkt\infty$-phase parent,
there are no walls.  The pictured connected spin excitations,
spanning $p_{\bkt1}+p_{\bkt\infty}-1=1$ block,
is the shortest that is possible for $\bkt2$ but impossible for
either parent.

The energy difference \eqref{eqn:energydiff} subtracts from the
free energy of diagram \eqref{eqn:2phase} the parent-diagram free energies.
We accomplish this with a product of vectors (lowercase Greek letters)
and matrices.  For this example, we get
\begin{equation}
a_{\bkt2}~\propto~
(\beta^\dag - \alpha^\dag)~{\cal A}~(\alpha - \beta) \rlap{\quad,}
\end{equation}
where $\alpha$ represents a diagram $S\hat S S\nonmag S$,
$\beta$ the diagram $S\hat S S\nonmag\wall S$, and $\cal A$ the
diagram $S\hat S S\hat S S$.
The duality operator, defined for vectors by $v^\dag= (Q v)^T$, with
$Q$ having $-1$ all along the antidiagonal, describes the
reversed diagram, {\it e.g.}, $\alpha^\dag= S\nonmag S \hat S S$,
with the clock directions also reversed.
Since the spin in an excitation can be rotated $60^\circ$ counterclockwise
($+$) or clockwise ($-$), both conditions must be accounted for.
The four entries of a matrix stand for the four
ways the two connected spins in a matrix diagram can be flipped:
\begin{equation}
\label{eqn:2x2}
\pmatrix{+- & ++ \cr -- & -+ \cr}\rlap{\quad.}
\end{equation}
The entries of a row vector are $(+~-)$, those of a column
vector $\left({-\atop+}\right)$, so that each contraction in a
matrix product sums over the possibilities for a single spin.
Each $2\times2$ matrix entry gives Boltzmann weights for connected
and disconnected combinations of the two constituent spins, as illustrated
in Figure \ref{fig:spins}a.

\def\topbox#1{\vtop{\vbox{}\hbox{#1}}}
\begin{figure}[htb]
$$
\vcenter{
\halign{#\ \hfil&\topbox{\epsfxsize0.3\hsize\epsfbox{#}}\quad\quad&%
#\ \hfil&\topbox{\epsfxsize0.3\hsize\epsfbox{#}}\cr
(a)&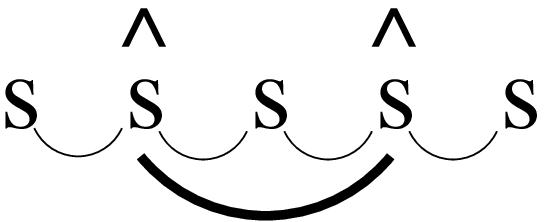&(c)&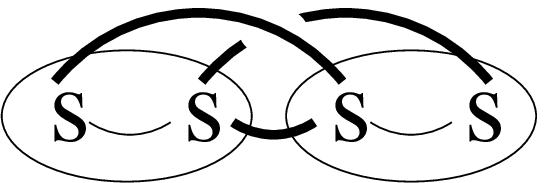\cr
\noalign{\vskip\baselineskip}
(b)&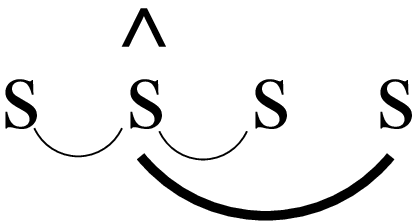&(d)&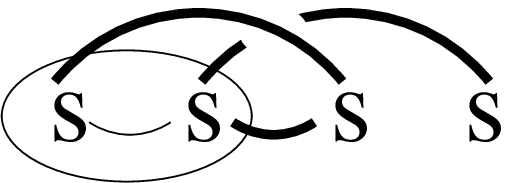\cr
}
}
$$
\caption{\label{fig:spins} A matrix element represents two flipped spins,
a vector element one.
Boldface bonds are counted at full strength in
the Boltzmann weights, while
each of the other bonds is counted in two different diagrams and so comes
in at half strength.
(a) A $2\times2$ matrix represents flipped
spins ($\hat S$) in the second and fourth planes.
(b) A (column) $2$-vector contracts with a $2\times2$ matrix to its left.
(c) A $4\times4$
matrix represents a flipped spin in one (and only one) of the first
two layers and in one (and only one) of the second two.
(d) A (column) $4$-vector contracts with a $4\times4$ matrix to its left.
(Adapted from Reference \cite{Seno}.)}
\end{figure}

In addition,
each matrix entry is a difference between the connected Boltzmann
factor and the disconnected factor, as specified by the
linked-cluster theorem, \eqref{eqn:linkedcluster}.
Vectors terminate the product (Figure \ref{fig:spins}b).
The following $2\times2$ matrices are required;
common factors of $q^t$ are omitted, since only the signs of the
matrix products in the zero-temperature limit matter.

{
\jot4\jot
\begin{eqnarray}
S\hat{S}S\hat{S}S\qquad &
{\cal A} =& r
\pmatrix{1 - r^{x} & r^{3x} - r^{4x} \cr
1 - r^{-2x} & 1 - r^{x} \cr}\\
S\nonmag\hat{S}S\hat{S}S\qquad &
{\cal B} =& 
\pmatrix{r^z - r^{x+z} & r^z(r^{3x} - r^{4x}) \cr
r^{{3\over2}-{z\over2}}(1 - r^{-2x}) & 
r^{{3\over2}-{z\over2}}(1 - r^{x}) \cr}\\
S\hat{S}\nonmag S\hat{S}S\qquad &
{\cal C} =& r^{z}
\pmatrix{r^{{3\over2}-{3z\over2}}(1 - r^{z}) & r^{{3\over2}+{3z\over2}}(
1 - r^{z})\cr 1 - r^{-2z} & 1 - r^{z} \cr}\\
S\nonmag\wall\hat{S}S\hat{S}S\qquad &
{\cal D} =& r^{{z\over2}}
\pmatrix{1 - r^{x} & r^{3x} - r^{4x} \cr
r^{{3\over2}(1-z)}(1 - r^{-2x}) & r^{{3\over2}(1-z)}(1 - r^{x}) \cr}\\
S\hat{S}\nonmag\wall S\hat{S}S\qquad &
{\cal E} =& r^{{z\over2}}
\pmatrix{r^{{3\over2}(1-z)}(1 - r^{2z})&r^{{3\over2}(1-z)}(r^{3z} - r^{2z})\cr
r^{3z} - r^{2z} & 1 - r^{2z} \cr}\\
S\hat{S}S\nonmag S\qquad &
\alpha =& r^{{1\over2}}
\pmatrix{r^{2z} \cr
r^{-z} \cr}\\
S\hat{S}S\nonmag\wall S\qquad &
\beta =& r^{{1\over2}}
\pmatrix{r^{z} \cr
r^{z} \cr}\\
\noalign{\noindent The following environments occur only when $L=3$:
\hfil\break\vskip0.25\jot}
S\nonmag\wall\hat{S}S\hat{S}\nonmag\wall S\qquad &
{\cal F} =& \pmatrix{r^{{1\over2}-{z\over2}}(1 - r^{x}) &
r^{3x+z-1}(1 - r^x)\cr
r^{2-2z}(1 - r^{-2x}) &
r^{{1\over2}-{z\over2}}(1 - r^x)\cr}\\
S\nonmag\wall\hat{S}S\hat{S}\nonmag S\qquad &
{\cal G} =& \pmatrix{r^{{1\over2}}(1 - r^x) & r^{-1+3x+{3z\over2}}(1 - r^x) \cr
r^{2-{3z\over2}}(1 - r^{-2x}) & r^{{1\over2}}(1 - r^x) \cr}\\
S\nonmag\hat{S}S\hat{S}\nonmag S\qquad &
{\cal H} =& \pmatrix{r^{{1\over2}+{z\over2}}(1 - r^x) &
r^{3x+2z-1}(1 - r^x) \cr r^{2-z}(1 - r^{-2x}) & r^{{1\over2}+{z\over2}}(
1 - r^x) \cr}
\end{eqnarray}
}

When $(p_a+p_b-1)\cdot L$ is even, there is no unique shortest leading-order
diagram on the model of \eqref{eqn:2phase}.  Rather, a family of such
diagrams with flipped spins every second layer except for one
pair of axially adjacent
flipped spins all span the requisite distance.  To account for
a single adjacent pair anywhere along the length of an excitation,
Seno {\it et al.} \cite{Seno} introduced $4\times4$ transfer matrices of
the form
\begin{equation}
\pmatrix{+0+0 & +0-0 & +00+ & +00- \cr 
         -0+0 & -0-0 & -00+ & -00- \cr
	 0++0 & 0+-0 & 0+0+ & 0+0- \cr
	 0-+0 & 0--0 & 0-0+ & 0-0- \cr} \rlap{\quad,}
\end{equation}
each entry of which considers four adjacent planes in which a spin
has rotated in the positive ($+$) or negative ($-$) clock direction, or
not rotated at all ($0$).  See Figure \ref{fig:spins}c.
The four entries of the upper-right quadrant
contain no connected spin excitations and so vanish.
End-cap 
vectors (Figure \ref{fig:spins}d) acount for the
final pair of planes, one of which will contain a spin flip.
The following matrices and end-cap vectors result
(again, the common factor of $q^t$ is omitted):

{
\jot4\jot
\def\quad{\hskip0.85em\relax}
\begin{eqnarray}
\fl
{\widehat{SS}\widehat{SS}}\quad&
A\!=\!
\pmatrix{r(1 - r^x) & r^{-{1\over2}}(r^{3x} - r^{4x}) & 0 & 0 \cr
r^{{5\over2}}(1 - r^{-2x}) & r(1 - r^{x}) & 0 & 0 \cr
r^{{1\over2}+x}(1 - r) & r^{4x}(r^{2} - 1) & r(1 - r^{x}) &
r^{-{1\over2}}(r^{3x} - r^{4x}) \cr
r^{-2x}(r^{2} - r^{3}) & r^{{1\over2}+x}(1 - r) &
r^{{5\over2}}(1 - r^{-2x}) & r(1 - r^{x}) \cr}\\
\fl
{\widehat{SS}\widehat{S\nonmag S}}\quad&
B\!=\!
\pmatrix{r^{{3\over2}-{z\over2}}(1 - r^{x}) & r^{z-{3\over2}}(r^{3x} -
r^{4x}) & 0 & 0 \cr
r^{3-{z\over2}}(1 - r^{-2x}) & r^{z}(1 - r^{x}) & 0 & 0 \cr
r^{{3z\over2}-x}(r - r^{2}) & r^{2x+3z}(r - r^{-1}) & r^{z}(1 - r^{z}) &
r^{5z\over 2} - r^{7z\over 2} \cr
r^{-{3z\over2}-x}(r^{{5\over2}} - r^{{7\over2}}) & r^{2x}(r^{-{1\over2}} -
r^{{1\over2}}) & r^{{3\over2}+z}(1 - r^{-2z}) & r^{{3\over2}-{z\over2}}
(1 - r^{z}) \cr}\\
\fl
{\widehat{SS}\nonmag\widehat{SS}}\quad&
C\!=\!
\pmatrix{r^{2z-1}(1 - r^{z}) & r^{{1\over2}}(r^{2z} - r^{3z}) & 0 & 0 \cr
r^{1\over2}(r^{2z} - 1) & r^2(r^{-z} - 1) & 0 & 0 \cr
r^{{1\over2}+z}(1 - r^{z}) & r^{2+2z}(r^{2z} - 1) & r^{2-z}(1 - r^{z}) &
r^{{1\over2}+2z}(1 - r^z) \cr
r^{z-1}(1 - r^z) & r^{{1\over2}+z}(1 - r^{z}) & r^{{1\over2}+
2z}(1 - r^{-2z}) & r^{2z-1}(1 - r^{z}) \cr}\\
\fl
{\widehat{SS}\widehat{S\nonmag\wall S}}\quad&
D\!=\!
\pmatrix{r^{{3\over2}-z}(1 - r^{x}) & r^{{z\over2}-{3\over2}}(r^{3x} -
r^{4x}) & 0 & 0 \cr
r^{3-z}(1 - r^{-2x}) & r^{z\over2}(1 - r^x) & 0 & 0 \cr
r^{-x}(r - r^{2}) & r^{{3z\over2}+2x}(r - r^{-1}) & r^{{z\over2}}(1 - r^{2z})
& r^{z}(r^{z} - 1) \cr
r^{-x}(r^{{5\over2}} - r^{{7\over2}}) & r^{2x+{3z\over2}}(r^{-{1\over2}} -
r^{{1\over2}}) &  r^{{3\over2}+{5z\over2}}(r^z - 1) &  r^{{3\over2}-
z}(1 - r^{2z}) \cr}\\
\fl
{\widehat{SS}\nonmag\wall\widehat{SS}}\quad&
E\!=\!
\pmatrix{r^{z-1}(1 - r^{2z})&r^{1\over2}(r^{z} - 1) & 0  & 0  \cr
r^{{1\over2}+3z}(r^{z} - 1) & r^{2-2z}(1 - r^{2z}) & 0 & 0 \cr
r^{{1\over2}+2z}(1 - r^{-z}) & r^{2-z}(1 - r^{-z}) & r^2(r^{-2z} - 1) & 
r^{1\over2}(r^z - 1) \cr
r^{2z - 1}(1 - r^{2z}) & r^{{1\over2}+2z}(1 - r^{-z}) & r^{{1\over2}+3z}(r^z
 - 1) & r^{z-1}(1 - r^{2z}) \cr}
\end{eqnarray}
}
{
\jot4\jot
\begin{eqnarray}
\widehat{SS}\nonmag S S\qquad &
a=\pmatrix{r^{2z-{1\over2}}\cr r^{1-z}\cr r^{1+z}\cr r^{z-{1\over2}}\cr}\\
\widehat{SS}\nonmag\wall S S\qquad & 
b=\pmatrix{r^{z-{1\over2}}\cr r^{1+z}\cr r^{1-z} \cr
r^{2z-{1\over2}} \cr}
\end{eqnarray}
}

A straightforward computer algorithm generates the relevant sequence of
matrices; as one example, for $L=4$,
\begin{equation}
a_{\bkt{23}} = (-a^\dag +b^\dag) A C A E A C A (a-b)
\rlap{\quad.}
\end{equation}
The programme
expands and symbolically determines the leading behaviour of $a$ in the
zero-temperature limit; if $a$ is negative, the child attains a
region of stability with respect to its parents.  For sufficiently
long chains of matrices, it was impractical to expand the matrix
products, and a numerical approach was substituted.
}

\section*{References}
\bibliographystyle{unsrt}
\bibliography{phases}

\end{document}